\documentclass[aps,pra,twocolumn]{revtex4-1}
\usepackage{graphicx}% Include figure files
\usepackage{dcolumn}% Align table columns on decimal point
\usepackage{amsmath}% add hypertext capabilities
\newcommand{\tn}[1]{\textnormal{#1}}
\newcommand{\OL}[1]{\overline{#1}}
\newcommand{\extravspace}{\vphantom{$^{^\tn{N}}$}}% insert extra vspace
\newcommand{\pt}[1]{\phantom{#1}} % shortcut for \phantom{}
\newcommand{\iso}[1]{$^{#1}$Ne} %\iso{20} will become $^{20}$Ne
\newcommand{\lra}{\leftrightarrow} % shortcut for \leftrightarrow

\begin{document}

\title{Heteronuclear collisions between laser-cooled metastable neon atoms\\
-\\
Phys. Rev. A \bf{86}, 022713 (2012)}
\author{Jan Sch\"utz}
\author{Thomas Feldker}  
\author{Holger John}
\author{Gerhard Birkl} 
\email{gerhard.birkl@physik.tu-darmstadt.de}
\affiliation{Institut f\"ur Angewandte Physik, Technische Universit\"at Darmstadt, Schlossgartenstra\ss{}e 7, D-64289 Darmstadt, Germany}
\date{\today}

\begin{abstract}
We investigate heteronuclear collisions in isotope mixtures of laser-cooled metastable ($^3P_2$) neon. Experiments are performed with spin-polarized atoms in a magnetic trap for all two-isotope combinations of the stable neon isotopes \iso{20}, \iso{21}, and \iso{22}. 
We determine the rate coefficients for heteronuclear ionizing collisions to $\beta_{21,20}=(3.9\pm2.7)\times 10^{-11}$\,cm$^3$/s, $\beta_{22,20}=(2.6\pm0.7)\times 10^{-11}$\,cm$^3$/s, and $\beta_{21,22}=(3.9\pm1.9)\times 10^{-11}$\,cm$^3$/s. We also study heteronuclear elastic collision processes and give upper bounds for heteronuclear thermal relaxation cross sections. This work significantly extends the limited available experimental data on heteronuclear ionizing collisions for laser-cooled atoms involving one or more rare gas atoms in a metastable state.
\end{abstract}

%\keywords{}

\maketitle

\section{Introduction \label{sec:intro}} 
Due to their high internal energy which surpasses the ionization energy of almost all neutral atomic and molecular collision partners, metastable rare-gas atoms ($Rg$*) prepared by laser cooling techniques \cite{CIGMA:12} present a unique class of atoms for the investigation of cold and ultracold collisions \cite{review_cold_collisions}. 
With lifetimes ranging from 14.73\,s (neon) \cite{Zinner:03} to 7870\,s (helium) \cite{Hodgman:09}, the first excited metastable states present effective ground states for applying standard laser cooling techniques. 
The resulting extremely low kinetic energy of the laser-cooled atoms ($\approx 10^{-10}$\,eV) stands in strong contrast to the high internal energy between 8\,eV (xenon) and 20\,eV (helium).
This offers unique conditions not accessible with samples of laser-cooled alkali-metal atoms.
In particular, the high internal energy of the metastable state allows for Penning (PI) and associative (AI) ionizing collisions,
\begin{eqnarray} \label{eq:penning}
  Rg^* + Rg^* &\rightarrow& Rg^* + Rg^+ + e^-\quad (\textnormal{PI}),\\	
  Rg^* + Rg^* &\rightarrow& Rg_2^+ + e^-\quad (\textnormal{AI}),	
\end{eqnarray}
to dominate losses in trapped $Rg$* samples at high densities. 
Since direct detection of the collision products (using electron multipliers) as well as of the remaining $Rg$* atoms is possible with high efficiency, detailed studies of homonuclear ionizing collisions have been carried out for all rare gas elements (see \cite{CIGMA:12} for an overview).
Additional motivation for these investigations arises from the fact, that favorable rate coefficients for inelastic but also for elastic collisions are essential for the achievement of quantum degeneracy which for $Rg$* atoms so far has only been demonstrated for $^{4}$He \cite{Robert:2001,Santos:2001,Tychkov:06,Dall:07} and $^{3}$He \cite{McNamara:06}.

For unpolarized $Rg$* samples, two-body loss rate coefficients for PI \cite{note1} between $\beta = 6 \times 10^{-11}\tn{cm}^3/\tn{s}$ for $^{132}$Xe \cite{Walhout:95} and $\beta = 1 \times 10^{-9}\tn{cm}^3/\tn{s}$ for $^{22}$Ne \cite{VanDrunen:08} have been measured.
Spin polarizing the atoms to a spin-stretched state may suppress PI: In a collision of atoms in spin-stretched states the total spin of the reactants is larger than the total spin of the products. Thus, if spin is conserved, PI collisions should be significantly reduced. In the case of He* in the $^3S_1$ state, a suppression of four orders of magnitude has been observed \cite{Tychkov:06,Stas:06}. For the heavier rare gases, however, the metastable state $^3P_2$ is formed by an exited $s$ electron and a $p^5$ core with orbital momentum $l=1$. This leads to an anisotropic quadrupole-quadrupole interaction between the colliding atoms that limits the amount of suppression \cite{Doery:98}. In the case of neon, suppression rates of 77 for \iso{20} and 83 for \iso{22}, respectively, have been determined \cite{CIGMA:12,VanDrunen:08}, but no suppression has been found for Kr* \cite{Katori:95} and Xe* \cite{Orzel:99}. 

In contrast to the case of {\it homonuclear} collisions, the number of experimental studies of cold {\it heteronuclear} ionizing collisions with at least one $Rg$* atom is very limited. To our knowledge, there exists only one investigation involving heteronuclear pairs of $Rg$* atoms, performed with metastable helium \cite{McNamara:07}, and only two experiments involving one $Rg$* atom in combination with a ground state alkali-metal atom: Ar* and Rb \cite{Busch:06} and He* and Rb \cite{Byron:10}. It is the purpose of the present work to extend the available experimental data by the investigation of heteronuclear collision between all possible combinations of the stable isotopes of neon. There exist two stable bosonic isotopes, \iso{20} (natural abundance: 90.48\%) and \iso{22} (9.25\%), and one fermionic isotope, \iso{21}, with an abundance of 0.27\% \cite{Rosman:98}. 
This work complements our previous measurements on homonuclear collisions for the bosonic isotopes \iso{20} and \iso{22} \cite{Spoden:05}.

In this paper we report on trapping of all three stable isotopes and all two-isotope mixtures of laser-cooled Ne* in a magnetic trap. We present measurements on the rates of heteronuclear ionizing collisions and heteronuclear thermal relaxation cross sections in isotope mixtures of spin-polarized Ne*. 
Specific emphasis has been put on the preparation of spin-polarized samples in order to have clear experimental conditions avoiding unambiguity arising from averaging over different internal state combinations during collisions.
We also address the question of whether the rates for heteronuclear collisional interactions are favorable for sympathetic cooling between different neon isotopes, as is the case for He* \cite{McNamara:06}. Sympathetic cooling of \iso{22}, which has a positive scattering length of  $a = 150\,a_0$, \cite{Spoden:05} using \iso{20} as a coolant might open a novel route to produce quantum degenerate bosonic Ne*. Sympathetic cooling of \iso{21}, using one of the bosonic isotopes as a coolant, could allow for the creation of a cold, dense sample of spin-polarized fermionic neon which could complement experiments with fermionic $^{3}$He for studying, e.g., the effect of Pauli blocking \cite{DeMarco:01} on Penning ionization for heavier rare-gas atoms \cite{Orzel:99}.

\subsection{Experimental setup \label{sec:setup}} 
In our experiment \cite{Zinner:03,Spoden:05}, neon atoms are excited to the metastable $^3P_2$ state in a dc discharge, then optically collimated, Zeeman decelerated, and captured in a magneto-optical trap. For laser cooling, the transition ${^3P_2} \lra {^3D_3}$ at 640.4\,nm is used. While the bosonic isotopes have no nuclear spin, \iso{21} has a nuclear spin of $I=3/2$ which causes a hyperfine structure of the atomic states. \iso{21} is trapped in the state with total angular momentum $F=7/2$ and the transition $F=7/2 \lra F'=9/2$ is used for laser cooling.
After magneto-optical trapping, the atoms are spin polarized to the $m_J=+2$ substate ($F=7/2, m_F=7/2$ in the case of \iso{21}) using circular polarized light and are transferred to an Ioffe-Pritchard magnetic trap (see \cite{Spoden:05} for details). As a last step of preparation, we apply one-dimensional Doppler cooling \cite{Schmidt:03} by irradiating the magnetically trapped atoms with circular polarized, red-detuned laser light along the symmetry axis of the trap.

The different laser fields for decelerating, cooling, optical trapping, and detecting the atoms of one isotope are derived from a single dye laser using acousto-optic frequency shifters. In order to trap \iso{21}, in addition to the cooling light, we superimpose ``repumping'' light ($F=5/2 \lra F'=7/2$) on all laser beams. This light pumps atoms that have been off-resonantly excited to the $F'=7/2$ state and have spontaneously decayed to the $F=5/2$ state back into the cooling cycle. In order to maximize the number of trapped \iso{21} atoms, we pump atoms that have been initially excited to the $F=3/2$ or $1/2$ states by the discharge source into the cooling cycle. For this purpose, we add laser light tuned between the closely spaced $F=3/2\lra F'=5/2$ and  $F=1/2\lra F'=3/2$ transitions in the optical collimation zone directly after the discharge source (see \cite{Feldker:11} for details). 

The isotope shift of the cooling transition is $(1625.9\pm0.15)$\,MHz for \iso{20} $\lra$ \iso{22} and $(1018.8\pm0.25)$\,MHz for \iso{20} $\lra$ \iso{21}, respectively \cite{Feldker:11}. 
In order to trap two isotopes simultaneously, we use two dye lasers with the frequency of each laser being stabilized to the cooling transition of the respective isotope and a non polarizing $50/50$ beam splitter to overlap their output light. Typical atom numbers of two-isotope mixtures of magnetically trapped atoms, compared to single isotope samples, are presented in Table~\ref{tab:numbers}. The atom numbers in two-isotope samples are  slightly lower than in the single isotope cases. The number of \iso{20} and \iso{22} atoms is limited due to inelastic collisions between trapped atoms, while the number of \iso{21} atoms is limited by the maximum available loading duration of 0.7\,s in our setup. After one-dimensional Doppler cooling, the mean temperatures are typically 300\,$\mu$K for single isotopes and 500\,$\mu$K for two-isotope mixtures.

For the bosonic isotopes, the number of trapped atoms is determined by absorption imaging. The images are taken 3\,ms after the trap has been switched off. Knowing the trapping potential, these images also allow for a determination of the ensemble temperatures from the widths of the atom clouds. In the case of isotope mixtures, the images are taken using light resonant with the cooling transition of one of the isotopes. Since the isotope shift of the cooling transition is at least 600\,MHz, the presence of another isotope does not influence the imaging process. 
For the given densities, absorption imaging is limited to atom samples of more than $10^6$ atoms. In order to determine the number of \iso{21} atoms, either fluorescence imaging or electronic detection are used. For fluorescence imaging of the magnetically trapped atoms, we illuminate the atoms with the magneto optical trap (MOT) beams for 100\,ms and collect the fluorescence light with the CCD camera. The signal is well pronounced even for small samples but does not reveal information about the temperature of the cloud. For electronic detection, we use a double micro channel plate (MCP) detector which is placed 13.9\,cm below and 2.7\,cm sideways with respect to the trap center. From the time-of-flight signal of Ne* released from the magnetic trap, the number of atoms as well as the temperature of the atom sample can be determined. To distinguish between different isotopes we push away the atoms of the unobserved isotope using slightly blue-detuned MOT light. We do not see any influence of this light on the remaining atoms of the observed isotope. 

\begin{table}  
  \caption{\label{tab:numbers}%
  Typical atom numbers of single- and two-isotope samples in the magnetic trap. After one-dimensional Doppler cooling, the mean temperatures are typically 300\,$\mu$K for single isotopes and 500\,$\mu$K for two-isotope mixtures.}
  \begin{ruledtabular}
  \begin{tabular}{crcr}
  	\multicolumn{2}{c}{Single-isotope samples} &
  	\multicolumn{2}{c}{Two-isotope mixtures}\\
    Isotope & Atoms & Isotopes & Atoms\\
    \hline 
    \extravspace
    \iso{20} &  $200\times10^6$ & \iso{20} + \iso{22} & $(170 + 50)\times10^6$\\
    \iso{21} &    $2\times10^6$ & \iso{20} + \iso{21} & $(120 +  1)\times10^6$\\
    \iso{22} &   $80\times10^6$ & \iso{22} + \iso{21} & $( 50 +  1)\times10^6$\\
  \end{tabular}
  \end{ruledtabular}
\end{table}

Since one-dimensional Doppler cooling is more efficient along the beam axis (i.e., the symmetry axis of the Ioffe-Pritchard trap) \cite{Schmidt:03}, we obtain ensembles not in three-dimensional thermal equilibrium after the preparation process. In each dimension, however, energy equipartition is fulfilled and the ensembles can be described by independent temperatures $T_{\tn{ax}}$ in axial and $T_{\tn{r}}$ in radial directions. A mean temperature defined as $\overline{T}=(T_{\tn{ax}}+2\,T_{\tn{r}})/3$ is used to describe the thermal energy of the atom sample. The absorption images are taken perpendicular to the symmetry axis of the trap and allow for a determination of $T_{\tn{ax}}$ and $T_{\tn{r}}$ simultaneously. From the time-of-flight signal on the MCP detector only $T_{\tn{r}}$ can be deduced directly since $T_{\tn{ax}}$ mainly influences the total number of atoms hitting the detector. However, by calibrating the MCP signals with corresponding fluorescence images, we can derive axial and radial temperature for electronic detection as well.

\section{Heteronuclear ionizing collisions \label{sec:collisions}} 
\begin{figure} 
  \includegraphics[width=\linewidth]{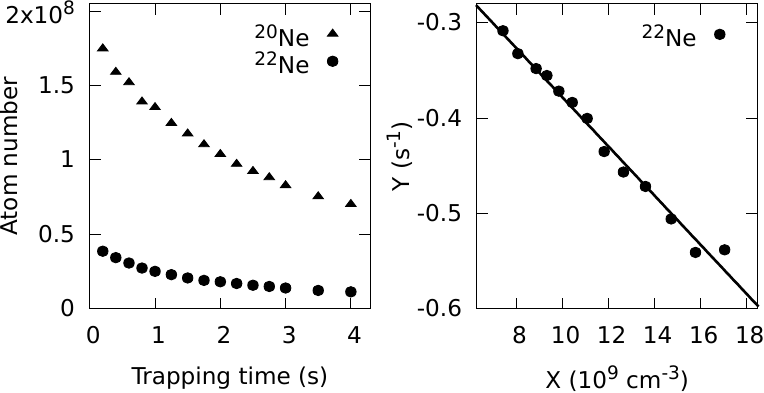}
  \caption{(Left) Number decay of spin-polarized \iso{20} and \iso{22} atoms in a two-isotope mixture in a  magnetic trap. Initially, $1.7\times10^8$ \iso{20} atoms at a mean density of $\overline{n}=2.5\times10^{10}$\,cm$^{-3}$ and $3.9\times10^7$ \iso{22} atoms at a mean density of $\overline{n}=3.8\times10^{9}$\,cm$^{-3}$ were loaded. Each data point is an average of six measurements. (Right) Atom number decay of \iso{22} presented according to Eq.~(\ref{eq:loss_integrated}) and a corresponding linear fit to the data. \label{fig:decay_2022}}
\end{figure}

We performed number decay measurements for all spin-polarized two-isotope mixtures in the magnetic trap for trapping times between 0.2\,s and 4\,s [see Fig.~\ref{fig:decay_2022} (left)]. After the trap is switched off, an individual image is recorded for one isotope and after a second identical repetition of the experimental run a corresponding image for the other isotope is taken.
For \iso{21}, in an additional run the time-of-flight signal on the MCP detector is recorded in order to gain full information about atom number $N$, $T_{\tn{ax}}$, and $T_{\tn{r}}$ for all isotopes.

From the number decay, we determine the rate coefficients of binary inelastic collisions. The decrease in atom number $N_i$ of isotope $^{i}$Ne originates from one-body losses ($\propto N_i$ with rate constant $\alpha_i$) as well as homonuclear ($\propto N_i^2$ with rate constant $\beta_{i}$) and heteronuclear ($\propto N_i N_j$ with rate constant $\beta_{i,j}$) two-body losses. We have no signature of higher order loss processes in our measurements. One-body losses are mainly caused by the 14.73\,s lifetime of the $^3P_2$ state and by background gas collisions. Two-body losses are almost entirely caused by ionizing collisions which we confirmed by simultaneously measuring the trap loss by absorption imaging and detecting the rate of ions escaping from the trap using the MCP in previous measurements for the bosonic isotopes \cite{Spoden:05}. Spin-changing two-body collisions play an insignificant role in two-body loss for our experimental conditions. We infer that this is also valid for \iso{21} since hyperfine state changing collisions are energetically suppressed due to the inverted hyperfine structure. 

The trap loss can be described by
\begin{eqnarray}\label{eq:loss}
\frac{d}{dt} N_i(t) &=& -\alpha_i N_i(t) - \beta_{i} \int d^3r \, n_i^2(\vec{r},t)\\
\nonumber
&& - \beta_{i,j} \int d^3r \, n_i(\vec{r},t) n_j(\vec{r},t)\\
\nonumber
&=& -\alpha_i N_i(t) - \beta_{i} \frac{N_i^2(t)}{V_{\text{eff},i}(t)} - \beta_{i,j} \frac{N_i(t)\,N_j(t)}{V_{\text{eff},i,j}(t)},
\end{eqnarray}
with densities $n_i(\vec{r},t)$ of isotope $^{i}$Ne. The trap potential can be calculated from the known geometry and the currents that are supplied to the coils of the trap which, together with the measured temperatures $T_{\tn{ax}}$ and $T_{\tn{r}}$, allows for a computation of the the effective volumes $V_{\text{eff},i}(t)$ and $V_{\text{eff},i,j}(t)$. 
For temperatures above 300\,$\mu$K the atoms are not fully localized in the harmonic region of the trap. Therefore, instead of assuming Gaussian density profiles, we perform the spatial integration numerically, also accounting for different $T_{\tn{ax}}$ and $T_{\tn{r}}$. 
All measurements were performed in the trap configuration that is used for Doppler cooling without further compression of the trap. Here, the trap is operated with an offset field of 2.6\,mT and the calculated vibrational trap frequencies for \iso{20} are $\omega_x = 2\pi\times80$\,s$^{-1}$ in axial and $\omega_r = 2\pi\times182$\,s$^{-1}$ in radial directions. 

We find that the effective volumes increase in time due to heating of the ensembles \cite{Spoden:05}. This heating is mainly caused by the higher rate of ionizing collisions at the trap center, involving atoms with an energy below average (cf. Sec. \ref{sec:elastic}). Explicit evaluation of Eq.~(\ref{eq:loss}) would require a detailed knowledge of the heating mechanisms, so we formally integrate Eq.~(\ref{eq:loss}) to obtain
\begin{multline} \label{eq:loss_integrated}
  \underset{\displaystyle Y}{\underbrace{\left[ 
	\left(\frac{1}{t}\ln{\frac{N_i(t)}{N_i(0)}}\right) 
	+ \beta_i\left(\frac{1}{t} \int^t_0 dt' \frac{N_i(t')}{V_{\text{eff},i}(t')} \right) 
	\right]}}\\ 
  = -\alpha_i  - \beta_{i,j} 
    \underset{\displaystyle X}{\underbrace{\left[ 
      \left(\frac{1}{t} \int^t_0 dt' \frac{N_j(t')}{V_{\text{eff},i,j}(t')} \right)
    \right]}}\,
\end{multline}
resulting in a linear relation $Y=-\alpha_i -\beta_{i,j}X$ with slope $\beta_{i,j}$ and offset $\alpha_i$. For \iso{20} and \iso{22} the homonuclear two-body loss coefficients $\beta_{20}=6.5(18)\times10^{-12}\,\mathrm{cm^3/s}$ and $\beta_{22}=1.2(3)\times10^{-11}\,\mathrm{cm^3/s}$ are known \cite{Spoden:05}. For \iso{21} homonuclear collisions can be neglected due to the small number of \iso{21} atoms relative to the numbers of \iso{20} and \iso{22} (see Table~\ref{tab:numbers}). All other quantities needed to calculate $X$ and $Y$ can be derived from experimental data by means of numerical integration.  

The number decay of \iso{20} and \iso{22} in a two-isotope mixture is shown in Fig.~\ref{fig:decay_2022} (left). The data points present averages of six measurements. The number decay of \iso{22} represented according to Eq.~(\ref{eq:loss_integrated}) is shown in Fig.~\ref{fig:decay_2022} (right). We observe the expected linear dependence and by a least-square fit we derive a one-body loss coefficient of $\alpha_{22}=(0.12\pm0.05)$\,s$^{-1}$ and a heteronuclear two-body loss coefficient of $\beta_{22,20}=(2.6\pm0.7)\times10^{-11}\,\mathrm{cm^3/s}$. The $\beta_{20,22}$ coefficient determined from the complementary number decay of \iso{20} in the same measurement agrees within the error margins  \cite{note2}. Similar analysis for the other isotope combinations yield $\beta_{21,20}=(3.9\pm2.7)\times10^{-11}\,\mathrm{cm^3/s}$ and $\beta_{21,22}=(3.9\pm1.9)\times10^{-11}\,\mathrm{cm^3/s}$, respectively. In the latter two cases, only the decay of \iso{21} can be analyzed since the effect of heteronuclear two-body loss induced by \iso{21} on one of the other isotopes is negligible due to the low number of trapped \iso{21} atoms.

The measured values for the rate coefficients of heteronuclear ionizing collisions together with the previously measured homonuclear rate coefficients for \iso{20} and \iso{22} are summarized in Table~\ref{tab:betas}. We find heteronuclear rate coefficients that are a factor of 2-6 larger than the homonuclear rate coefficients. A larger rate coefficient for heteronuclear collisions has also been observed in the corresponding measurements in He* \cite{McNamara:07}.

The quoted uncertainties are the combined statistical and systematic uncertainties. They are
dominated by systematic uncertainties in atom density, depending on atom number and effective volume. The evaluation procedure according to Eq.~(\ref{eq:loss_integrated}) is especially sensitive to the initial atom number. 
Other, also included but less dominant contributions to the uncertainties are given by the uncertainties of the homonuclear two-body loss coefficients. For \iso{21}, additional uncertainties arise due to the indirect measurement of $T_{\tn{ax}}$ and larger statistical uncertainties because of the low atom number.

\begin{table}
  \caption{Summary of heteronuclear ionizing collision rate coefficients of spin-polarized Ne* complemented by the homonuclear coefficients, taken from Spoden \textit{et al.}~\cite{Spoden:05}. \label{tab:betas}}
  \begin{ruledtabular}
  \begin{tabular}{ll}
     &  $\beta\ (10^{-11}\,\mathrm{cm^3/s})$ \\
	\hline
	Heteronuclear rate coefficients:\\
	\quad$\beta_{21,20}$ & $3.9 \pm 2.7$ \\
	\quad$\beta_{22,20}$ & $2.6 \pm 0.7$ \\
	\quad$\beta_{21,22}$ & $3.9 \pm 1.9$ \\
	Homonuclear rate coefficients \cite{Spoden:05}:\\
	\quad$\beta_{20}$ & $0.65 \pm 0.18$ \\
	\quad$\beta_{22}$ & $1.2 \pm 0.3$ \\
  \end{tabular}
  \end{ruledtabular}
\end{table}	

\section{Heteronuclear elastic collisions \label{sec:elastic}}
We also analyzed the heteronuclear elastic collision rates for magnetically trapped spin-polarized Ne*. In addition to the importance for completing the available database for metastable neon atoms, efficient sympathetic cooling between different isotopes requires a large ratio of elastic to inelastic collision rates. Therefore, we studied the heteronuclear elastic collision rates between isotope $^{i}$Ne and $^{j}$Ne by determining internuclear thermal relaxation rate constants $\gamma_{\text{rel},i,j}$, defined by
\begin{equation}\label{eq:gamma_rel}
	\frac{d}{dt} \overline{T}_i = \gamma_{\text{rel},i,j} \left( \overline{T}_j-\overline{T}_i\right)\,. 	
\end{equation} 
Since $\gamma_{\text{rel},i,j}$ depends on the densities and temperatures of the ensembles involved, it is useful to define an effective relaxation cross section
\begin{equation}\label{eq:sigma_rel}
	\sigma_{\text{rel},i,j} = \gamma_{\text{rel},i,j} \, \frac{V_{\text{eff},i,j}}{N_j \bar{v}}\,,
\end{equation}
where $\bar{v} = \sqrt{16 k_B \overline{T} /(\pi m)}$ is the average relative velocity of the colliding atoms.

We prepare two-isotope mixtures in the magnetic trap with a temperature difference between the isotopes by applying one-dimensional Doppler cooling with different efficiencies to the two isotopes. For all isotope combinations, we find that heteronuclear thermal relaxation is outbalanced by intrinsic heating caused by ionizing collisions: Since ionizing collisions occur preferentially at the center of the trap, where the density is highest and the energy of the atoms is lowest, these collisions remove atoms with below average energy. This leads to an effective heating of the remaining atoms (see also \cite{Soeding:98,Spoden:05}). The heating rate per ionizing collision depends on the mean energy $\overline{E}_C$ carried away by the collision products relative to the mean energy $\overline{E}_T$ of all atoms in the trap. The heating rate is given by \cite{Soeding:98}
\begin{equation}\label{eq:heating}
	\frac{d}{dt} \OL{T}_i =   c_{i,j}\, \beta_{i,j}\, \frac{N_j}{V_{\text{eff},i,j}}\, \OL{T}_i \,,
\end{equation}
with $c_{i,j}=1-\OL{E}_{C}/\OL{E}_{T}$. We calculate $c_{i,j}$ assuming a velocity-independent cross section and Gaussian density distributions. For homonuclear collisions $c_i=1/4$; for heteronuclear collisions $c_{i,j}$ depends on the temperatures of the two isotopes and is limited to $0\leq c_{i,j}\leq1/2$.

The time evolution of $\OL{T}_i$, thus can be described by
\begin{eqnarray}\label{eq:temperature}
	\nonumber
	\frac{d}{dt} \OL{T}_i &=& \gamma_{\text{rel},i,j} (\OL{T}_j-\OL{T}_i) + c_{i,j}\, \beta_{i,j}\, \frac{N_j}{V_{\text{eff},i,j}}\, \OL{T}_i\\
	&& + \frac{1}{4}\,\beta_i\, \frac{N_i}{V_{\text{eff},i}}\, \OL{T}_i + C\,,
\end{eqnarray}
with an additional constant heating rate $C$ accounting for density-independent heating effects. Again, we can formally integrate Eq. (\ref{eq:temperature}) to obtain
\begin{flalign}\label{eq:temperature_integrated}
	\left[\ln{\frac{\OL{T}_i(t)}{\OL{T}_i(0)}} + \int^t_0 dt' \left(\frac{\beta_{i,j}\,c_{i,j}(t')\,N_j(t')}{V_{\text{eff},i,j}(t')}
	+ \frac{\beta_i\,N_i(t')}{4 V_{\text{eff},i}(t')}\right. \right. \nonumber\\
	\underbrace{\hspace*{6cm}\left.\left. + \frac{C}{\OL{T}_i(t')}\right)\right]}_{\displaystyle Y'} \nonumber\\
= \sigma_{\text{rel},i,j} \underbrace{\left\{\int^t_0 dt' \left[ \frac{N_j(t') \bar{v}(t')}{V_{\text{eff},i,j}(t')} \left(\frac{\OL{T}_j(t')}{\OL{T}_i(t')} -1 \right) \right]\right\}}_{\displaystyle X'}\,,
\end{flalign}
resulting in a linear relation $Y'=\sigma_{\text{rel},i,j}X'$ with the slope given by $\sigma_{\text{rel},i,j}$. Again, all quantities needed to calculate $X'$ and $Y'$ can be derived from experimental data by means of numerical integration.

\begin{figure}
  \centering
  \includegraphics[width=\linewidth]{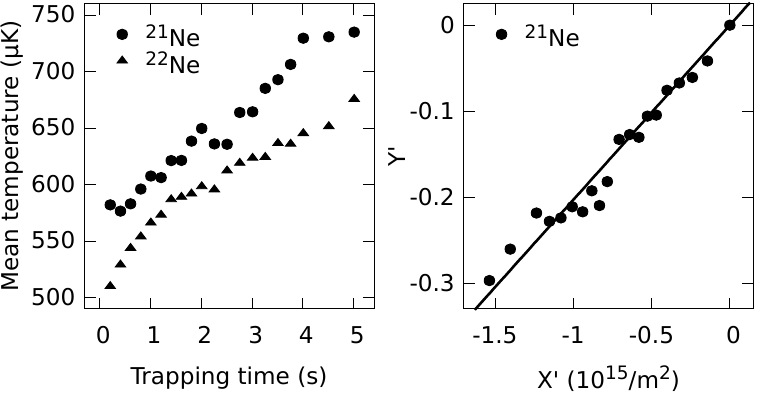}
  \caption{(Left) Time evolution of the mean temperature $\OL{T}$ of spin-polarized \iso{21} and \iso{22} atoms in a two-isotope mixture. The data points present averages of six measurements. Initially, $7\times10^{5}$ \iso{21} atoms with $T_{\tn{ax}}=330\,\mu\tn{K}$ and $T_{\tn{r}}=710\,\mu\tn{K}$ and $4.7\times10^{7}$ \iso{22} atoms with $T_{\tn{ax}}=360\,\mu\tn{K}$ and $T_{\tn{r}}=590\,\mu\tn{K}$ were trapped. (Right) Time evolution of $\OL{T}_{21}$ presented according to Eq.~(\ref{eq:temperature_integrated}) and corresponding linear fit to the data for determining the upper bound of $\sigma_{\text{rel},21,22}$. \label{fig:T_2122}}
\end{figure}

We performed relaxation cross-section measurements for all two-isotope combinations. 
As an example, the time evolution of the dimensionally averaged temperatures $\OL{T}$ of spin-polarized \iso{21} and \iso{22} in a two-isotope mixture and the representation of $\OL{T}_{21}$ according to Eq.~(\ref{eq:temperature_integrated}) are shown in Fig.~\ref{fig:T_2122}. 
All measurements reveal the systematic problem that the time evolution of the temperature in all cases is dominated by intrinsic heating: The temperature for all isotopes is increasing during the trapping time and thermal relaxation is not sufficiently efficient to equilibrate the temperatures of the different isotopes. 
Additionally, the large uncertainties in $\beta_{i,j}$ (and thus in the heating rates) and in density lead to huge uncertainties in the extracted relaxation cross sections. As a consequence, only upper bounds for the values of $\sigma_{\text{rel},i,j}$ can be determined. 

For the data presented in Fig.~\ref{fig:T_2122} (right), we observe the expected linear dependence and by a least-square fit we determine $\sigma_{\text{rel},21,22}(\approx600\,\mu\tn{K})\lesssim 20\times10^{-17}\,\tn{m}^2$. Homonuclear collisions of \iso{21} and effects of \iso{21} on \iso{22} can be neglected due to the low number of \iso{21} atoms. The time evolution of $\OL{T}_{22}$ is used to determine the constant heating rate $C$. A respective analysis for a \iso{20}-\iso{21} mixture yields $\sigma_{\text{rel},21,20}(\approx500\,\mu\tn{K}) \lesssim 7\times10^{-17}\,\tn{m}^2$. 
In case of \iso{20}-\iso{22} mixtures, the time evolution of $\OL{T}_{20}$ as well as $\OL{T}_{22}$ can be fitted. Analysis of $\OL{T}_{20}$ yields $\sigma_{\text{rel},20,22}(\approx750\,\mu\tn{K}) \lesssim 4\times10^{-17}\,\tn{m}^2$. 
By fitting $\OL{T}_{22}$ we get a value for $\sigma_{\text{rel},22,20}$ consistent with this upper bound but exhibiting even stronger heating effects not included in our model based on Eq.~(\ref{eq:heating}). This might indicate that the assumption of a velocity-independent cross section is not valid or that the differences in $T_{\tn{ax}}$ and $T_{\tn{r}}$ need to be taken into account in a more sophisticated fashion  \cite{note2}. 

The resulting upper bounds of $\sigma_{\text{rel},i,j}$ for all two-isotope mixtures are summarized in Table~\ref{tab:sigmas} and compared to previously measured homonuclear thermal relaxation cross sections of the bosonic isotopes \cite{Spoden:05}. The values for homonuclear and the upper bounds for heteronuclear cross sections have the same order of magnitude, although significantly smaller values for the heteronuclear cross sections could be possible. In any case, the resulting upper bounds exclude the possibility that strongly enhanced heteronuclear thermalization could lead to efficient two-isotope evaporative cooling.
\begin{table}
  \caption{Summary of upper bounds for heteronuclear thermal relaxation cross sections of spin-polarized Ne* compared to homonuclear cross sections, taken from Spoden \textit{et al.}~\cite{Spoden:05}. \label{tab:sigmas}}
  \begin{ruledtabular}
  \begin{tabular}{ll}
     &  $\sigma_{\text{rel}} (10^{-17}\,\tn{m}^2)$ \\
	\hline
	Heteronuclear relaxation cross sections:\\
	\quad $\sigma_{\text{rel},21,20} (\approx500\,\mu\tn{K})$   &   $\lesssim 7$ \\
	\quad $\sigma_{\text{rel},20,22} (\approx750\,\mu\tn{K})$   &   $\lesssim 4$ \\
	\quad $\sigma_{\text{rel},21,22} (\approx600\,\mu\tn{K})$   &   $\lesssim 20$ \\
	Homonuclear relaxation cross sections \cite{Spoden:05}:\\
	\quad $\sigma_{\text{rel},20} (550\,\mu\tn{K})$   &   $2.8\pm0.7$ \\
	\quad $\sigma_{\text{rel},22} (550\,\mu\tn{K})$   &   $\pt{.}13\pm3$ \\
  \end{tabular}
  \end{ruledtabular}
\end{table}	
\section{Conclusion \label{sec:conclusion}}
We measured the rate coefficients for heteronuclear ionizing collisions of spin-polarized Ne* and determined upper bounds for the thermal relaxation cross sections. We found that the heteronuclear ionizing collision rate coefficients are a factor of 2 to 6 larger than their homonuclear counterparts, while the heteronuclear and homonuclear thermal relaxation cross sections are expected to have comparable values. This shows that the heteronuclear collisional interaction properties are not favorable for hetero nuclear sympathetic cooling and that the ``good-to-bad'' ratio in the collisional properties is worse in the heteronuclear than in the homonuclear cases.
\begin{acknowledgments}
This work was supported in part by the German Research Foundation (DFG) (Contract No. BI 647/3-1) and by the European Science Foundation (ESF) within the Collaborative Research Project CIGMA of the EUROCORES program EuroQUAM. 
\end{acknowledgments}

\bibliographystyle{apsrev}		
\bibliography{ne_multi_isotopes}

\begin{thebibliography}{26}
\expandafter\ifx\csname natexlab\endcsname\relax\def\natexlab#1{#1}\fi
\expandafter\ifx\csname bibnamefont\endcsname\relax
  \def\bibnamefont#1{#1}\fi
\expandafter\ifx\csname bibfnamefont\endcsname\relax
  \def\bibfnamefont#1{#1}\fi
\expandafter\ifx\csname citenamefont\endcsname\relax
  \def\citenamefont#1{#1}\fi
\expandafter\ifx\csname url\endcsname\relax
  \def\url#1{\texttt{#1}}\fi
\expandafter\ifx\csname urlprefix\endcsname\relax\def\urlprefix{URL }\fi
\providecommand{\bibinfo}[2]{#2}
\providecommand{\eprint}[2][]{\url{#2}}

\bibitem[{\citenamefont{Vassen et~al.}(2012)\citenamefont{Vassen,
  Cohen-Tannoudji, Leduc, Boiron, Westbrook, Truscott, Baldwin, Birkl, Cancio,
  and Trippenbach}}]{CIGMA:12}
\bibinfo{author}{\bibfnamefont{W.}~\bibnamefont{Vassen}},
  \bibinfo{author}{\bibfnamefont{C.}~\bibnamefont{Cohen-Tannoudji}},
  \bibinfo{author}{\bibfnamefont{M.}~\bibnamefont{Leduc}},
  \bibinfo{author}{\bibfnamefont{D.}~\bibnamefont{Boiron}},
  \bibinfo{author}{\bibfnamefont{C.}~\bibnamefont{Westbrook}},
  \bibinfo{author}{\bibfnamefont{A.}~\bibnamefont{Truscott}},
  \bibinfo{author}{\bibfnamefont{K.}~\bibnamefont{Baldwin}},
  \bibinfo{author}{\bibfnamefont{G.}~\bibnamefont{Birkl}},
  \bibinfo{author}{\bibfnamefont{P.}~\bibnamefont{Cancio}}, \bibnamefont{and}
  \bibinfo{author}{\bibfnamefont{M.}~\bibnamefont{Trippenbach}},
  \bibinfo{journal}{Rev. Mod. Phys.} \textbf{\bibinfo{volume}{84}},
  \bibinfo{pages}{175} (\bibinfo{year}{2012}).

\bibitem[{\citenamefont{Weiner et~al.}(1999)\citenamefont{Weiner, Bagnato,
  Zilio, and Julienne}}]{review_cold_collisions}
\bibinfo{author}{\bibfnamefont{J.}~\bibnamefont{Weiner}},
  \bibinfo{author}{\bibfnamefont{V.~S.} \bibnamefont{Bagnato}},
  \bibinfo{author}{\bibfnamefont{S.}~\bibnamefont{Zilio}}, \bibnamefont{and}
  \bibinfo{author}{\bibfnamefont{P.}~\bibnamefont{Julienne}},
  \bibinfo{journal}{Rev. Mod. Phys.} \textbf{\bibinfo{volume}{71}},
  \bibinfo{pages}{1} (\bibinfo{year}{1999}).

\bibitem[{\citenamefont{Zinner et~al.}(2003)\citenamefont{Zinner, Spoden,
  Kraemer, Birkl, and Ertmer}}]{Zinner:03}
\bibinfo{author}{\bibfnamefont{M.}~\bibnamefont{Zinner}},
  \bibinfo{author}{\bibfnamefont{P.}~\bibnamefont{Spoden}},
  \bibinfo{author}{\bibfnamefont{T.}~\bibnamefont{Kraemer}},
  \bibinfo{author}{\bibfnamefont{G.}~\bibnamefont{Birkl}}, \bibnamefont{and}
  \bibinfo{author}{\bibfnamefont{W.}~\bibnamefont{Ertmer}},
  \bibinfo{journal}{Phys. Rev. A} \textbf{\bibinfo{volume}{67}},
  \bibinfo{pages}{010501} (\bibinfo{year}{2003}).

\bibitem[{\citenamefont{Hodgman et~al.}(2009)\citenamefont{Hodgman, Dall,
  Byron, Baldwin, Buckman, and Truscott}}]{Hodgman:09}
\bibinfo{author}{\bibfnamefont{S.~S.} \bibnamefont{Hodgman}},
  \bibinfo{author}{\bibfnamefont{R.~G.} \bibnamefont{Dall}},
  \bibinfo{author}{\bibfnamefont{L.~J.} \bibnamefont{Byron}},
  \bibinfo{author}{\bibfnamefont{K.~G.~H.} \bibnamefont{Baldwin}},
  \bibinfo{author}{\bibfnamefont{S.~J.} \bibnamefont{Buckman}},
  \bibnamefont{and} \bibinfo{author}{\bibfnamefont{A.~G.}
  \bibnamefont{Truscott}}, \bibinfo{journal}{Phys. Rev. Lett.}
  \textbf{\bibinfo{volume}{103}}, \bibinfo{pages}{053002}
  (\bibinfo{year}{2009}).

\bibitem[{\citenamefont{Robert et~al.}(2001)\citenamefont{Robert, Sirjean,
  Browaeys, Poupard, Nowak, Boiron, Westbrook, and Aspect}}]{Robert:2001}
\bibinfo{author}{\bibfnamefont{A.}~\bibnamefont{Robert}},
  \bibinfo{author}{\bibfnamefont{O.}~\bibnamefont{Sirjean}},
  \bibinfo{author}{\bibfnamefont{A.}~\bibnamefont{Browaeys}},
  \bibinfo{author}{\bibfnamefont{J.}~\bibnamefont{Poupard}},
  \bibinfo{author}{\bibfnamefont{S.}~\bibnamefont{Nowak}},
  \bibinfo{author}{\bibfnamefont{D.}~\bibnamefont{Boiron}},
  \bibinfo{author}{\bibfnamefont{C.~I.} \bibnamefont{Westbrook}},
  \bibnamefont{and} \bibinfo{author}{\bibfnamefont{A.}~\bibnamefont{Aspect}},
  \bibinfo{journal}{Science} \textbf{\bibinfo{volume}{292}},
  \bibinfo{pages}{461} (\bibinfo{year}{2001}).

\bibitem[{\citenamefont{{Pereira Dos Santos}
  et~al.}(2001)\citenamefont{{Pereira Dos Santos}, L\'{e}onard, Wang, Barrelet,
  Perales, Rasel, Unnikrishnan, Leduc, and Cohen-Tannoudji}}]{Santos:2001}
\bibinfo{author}{\bibfnamefont{F.}~\bibnamefont{{Pereira Dos Santos}}},
  \bibinfo{author}{\bibfnamefont{J.}~\bibnamefont{L\'{e}onard}},
  \bibinfo{author}{\bibfnamefont{J.}~\bibnamefont{Wang}},
  \bibinfo{author}{\bibfnamefont{C.~J.} \bibnamefont{Barrelet}},
  \bibinfo{author}{\bibfnamefont{F.}~\bibnamefont{Perales}},
  \bibinfo{author}{\bibfnamefont{E.}~\bibnamefont{Rasel}},
  \bibinfo{author}{\bibfnamefont{C.~S.} \bibnamefont{Unnikrishnan}},
  \bibinfo{author}{\bibfnamefont{M.}~\bibnamefont{Leduc}}, \bibnamefont{and}
  \bibinfo{author}{\bibfnamefont{C.}~\bibnamefont{Cohen-Tannoudji}},
  \bibinfo{journal}{Phys. Rev. Lett.} \textbf{\bibinfo{volume}{86}},
  \bibinfo{pages}{3459} (\bibinfo{year}{2001}).

\bibitem[{\citenamefont{Tychkov et~al.}(2006)\citenamefont{Tychkov, Jeltes,
  McNamara, Tol, Herschbach, Hogervorst, and Vassen}}]{Tychkov:06}
\bibinfo{author}{\bibfnamefont{A.~S.} \bibnamefont{Tychkov}},
  \bibinfo{author}{\bibfnamefont{T.}~\bibnamefont{Jeltes}},
  \bibinfo{author}{\bibfnamefont{J.~M.} \bibnamefont{McNamara}},
  \bibinfo{author}{\bibfnamefont{P.~J.~J.} \bibnamefont{Tol}},
  \bibinfo{author}{\bibfnamefont{N.}~\bibnamefont{Herschbach}},
  \bibinfo{author}{\bibfnamefont{W.}~\bibnamefont{Hogervorst}},
  \bibnamefont{and} \bibinfo{author}{\bibfnamefont{W.}~\bibnamefont{Vassen}},
  \bibinfo{journal}{Phys. Rev. A} \textbf{\bibinfo{volume}{73}},
  \bibinfo{pages}{031603(R)} (\bibinfo{year}{2006}).

\bibitem[{\citenamefont{Dall and Truscott}(2007)}]{Dall:07}
\bibinfo{author}{\bibfnamefont{R.}~\bibnamefont{Dall}} \bibnamefont{and}
  \bibinfo{author}{\bibfnamefont{A.}~\bibnamefont{Truscott}},
  \bibinfo{journal}{Opt. Commun.} \textbf{\bibinfo{volume}{270}},
  \bibinfo{pages}{255} (\bibinfo{year}{2007}).

\bibitem[{\citenamefont{McNamara et~al.}(2006)\citenamefont{McNamara, Jeltes,
  Tychkov, Hogervorst, and Vassen}}]{McNamara:06}
\bibinfo{author}{\bibfnamefont{J.~M.} \bibnamefont{McNamara}},
  \bibinfo{author}{\bibfnamefont{T.}~\bibnamefont{Jeltes}},
  \bibinfo{author}{\bibfnamefont{A.~S.} \bibnamefont{Tychkov}},
  \bibinfo{author}{\bibfnamefont{W.}~\bibnamefont{Hogervorst}},
  \bibnamefont{and} \bibinfo{author}{\bibfnamefont{W.}~\bibnamefont{Vassen}},
  \bibinfo{journal}{Phys. Rev. Lett.} \textbf{\bibinfo{volume}{97}},
  \bibinfo{pages}{080404} (\bibinfo{year}{2006}).

\bibitem[{not({\natexlab{a}})}]{note1}
\bibinfo{note}{All experiments able to discriminate between PI and AI show a
  significant predominance of PI over AI \cite{CIGMA:12}, and we refer to the
  sum of both loss channels as PI for the remainder of this article.}

\bibitem[{\citenamefont{Walhout et~al.}(1995)\citenamefont{Walhout, Sterr,
  Orzel, Hoogerland, and Rolston}}]{Walhout:95}
\bibinfo{author}{\bibfnamefont{M.}~\bibnamefont{Walhout}},
  \bibinfo{author}{\bibfnamefont{U.}~\bibnamefont{Sterr}},
  \bibinfo{author}{\bibfnamefont{C.}~\bibnamefont{Orzel}},
  \bibinfo{author}{\bibfnamefont{M.}~\bibnamefont{Hoogerland}},
  \bibnamefont{and} \bibinfo{author}{\bibfnamefont{S.~L.}
  \bibnamefont{Rolston}}, \bibinfo{journal}{Phys. Rev. Lett.}
  \textbf{\bibinfo{volume}{74}}, \bibinfo{pages}{506} (\bibinfo{year}{1995}).

\bibitem[{\citenamefont{van Drunen}(2008)}]{VanDrunen:08}
\bibinfo{author}{\bibfnamefont{W.~J.} \bibnamefont{van Drunen}},
  \bibinfo{type}{{Ph.D. thesis}}, \bibinfo{school}{Technische Universit\"{a}t
  Darmstadt} (\bibinfo{year}{2008}).

\bibitem[{\citenamefont{Stas et~al.}(2006)\citenamefont{Stas, McNamara,
  Hogervorst, and Vassen}}]{Stas:06}
\bibinfo{author}{\bibfnamefont{R.~J.~W.} \bibnamefont{Stas}},
  \bibinfo{author}{\bibfnamefont{J.~M.} \bibnamefont{McNamara}},
  \bibinfo{author}{\bibfnamefont{W.}~\bibnamefont{Hogervorst}},
  \bibnamefont{and} \bibinfo{author}{\bibfnamefont{W.}~\bibnamefont{Vassen}},
  \bibinfo{journal}{Phys. Rev. A} \textbf{\bibinfo{volume}{73}},
  \bibinfo{pages}{032713} (\bibinfo{year}{2006}).

\bibitem[{\citenamefont{Doery et~al.}(1998)\citenamefont{Doery, Vredenbregt,
  {Op de Beek}, Beijerinck, and Verhaar}}]{Doery:98}
\bibinfo{author}{\bibfnamefont{M.~R.} \bibnamefont{Doery}},
  \bibinfo{author}{\bibfnamefont{E.~J.~D.} \bibnamefont{Vredenbregt}},
  \bibinfo{author}{\bibfnamefont{S.~S.} \bibnamefont{{Op de Beek}}},
  \bibinfo{author}{\bibfnamefont{H.~C.~W.} \bibnamefont{Beijerinck}},
  \bibnamefont{and} \bibinfo{author}{\bibfnamefont{B.~J.}
  \bibnamefont{Verhaar}}, \bibinfo{journal}{Phys. Rev. A}
  \textbf{\bibinfo{volume}{58}}, \bibinfo{pages}{3673} (\bibinfo{year}{1998}).

\bibitem[{\citenamefont{Katori et~al.}(1995)\citenamefont{Katori, Kunugita, and
  Ido}}]{Katori:95}
\bibinfo{author}{\bibfnamefont{H.}~\bibnamefont{Katori}},
  \bibinfo{author}{\bibfnamefont{H.}~\bibnamefont{Kunugita}}, \bibnamefont{and}
  \bibinfo{author}{\bibfnamefont{T.}~\bibnamefont{Ido}},
  \bibinfo{journal}{Phys. Rev. A} \textbf{\bibinfo{volume}{52}},
  \bibinfo{pages}{R4324} (\bibinfo{year}{1995}).

\bibitem[{\citenamefont{Orzel et~al.}(1999)\citenamefont{Orzel, Walhout, Sterr,
  Julienne, and Rolston}}]{Orzel:99}
\bibinfo{author}{\bibfnamefont{C.}~\bibnamefont{Orzel}},
  \bibinfo{author}{\bibfnamefont{M.}~\bibnamefont{Walhout}},
  \bibinfo{author}{\bibfnamefont{U.}~\bibnamefont{Sterr}},
  \bibinfo{author}{\bibfnamefont{P.~S.} \bibnamefont{Julienne}},
  \bibnamefont{and} \bibinfo{author}{\bibfnamefont{S.~L.}
  \bibnamefont{Rolston}}, \bibinfo{journal}{Phys. Rev. A}
  \textbf{\bibinfo{volume}{59}}, \bibinfo{pages}{1926} (\bibinfo{year}{1999}).

\bibitem[{\citenamefont{McNamara et~al.}(2007)\citenamefont{McNamara, Stas,
  Hogervorst, and Vassen}}]{McNamara:07}
\bibinfo{author}{\bibfnamefont{J.~M.} \bibnamefont{McNamara}},
  \bibinfo{author}{\bibfnamefont{R.~J.~W.} \bibnamefont{Stas}},
  \bibinfo{author}{\bibfnamefont{W.}~\bibnamefont{Hogervorst}},
  \bibnamefont{and} \bibinfo{author}{\bibfnamefont{W.}~\bibnamefont{Vassen}},
  \bibinfo{journal}{Phys. Rev. A} \textbf{\bibinfo{volume}{75}},
  \bibinfo{pages}{062715} (\bibinfo{year}{2007}).

\bibitem[{\citenamefont{Busch et~al.}(2006)\citenamefont{Busch, Shaffer, Ahmed,
  and Sukenik}}]{Busch:06}
\bibinfo{author}{\bibfnamefont{H.~C.} \bibnamefont{Busch}},
  \bibinfo{author}{\bibfnamefont{M.~K.} \bibnamefont{Shaffer}},
  \bibinfo{author}{\bibfnamefont{E.~M.} \bibnamefont{Ahmed}}, \bibnamefont{and}
  \bibinfo{author}{\bibfnamefont{C.~I.} \bibnamefont{Sukenik}},
  \bibinfo{journal}{Phys. Rev. A} \textbf{\bibinfo{volume}{73}},
  \bibinfo{pages}{023406} (\bibinfo{year}{2006}).

\bibitem[{\citenamefont{Byron et~al.}(2010)\citenamefont{Byron, Dall, Rugway,
  and Truscott}}]{Byron:10}
\bibinfo{author}{\bibfnamefont{L.~J.} \bibnamefont{Byron}},
  \bibinfo{author}{\bibfnamefont{R.~G.} \bibnamefont{Dall}},
  \bibinfo{author}{\bibfnamefont{W.}~\bibnamefont{Rugway}}, \bibnamefont{and}
  \bibinfo{author}{\bibfnamefont{A.~G.} \bibnamefont{Truscott}},
  \bibinfo{journal}{New Journal of Physics} \textbf{\bibinfo{volume}{12}},
  \bibinfo{pages}{013004} (\bibinfo{year}{2010}).

\bibitem[{\citenamefont{Rosman and Taylor}(1998)}]{Rosman:98}
\bibinfo{author}{\bibfnamefont{K.~J.~R.} \bibnamefont{Rosman}}
  \bibnamefont{and} \bibinfo{author}{\bibfnamefont{P.~D.~P.}
  \bibnamefont{Taylor}}, \bibinfo{journal}{J. Phys. Chem. Ref. Data}
  \textbf{\bibinfo{volume}{27}}, \bibinfo{pages}{1275} (\bibinfo{year}{1998}).

\bibitem[{\citenamefont{Spoden et~al.}(2005)\citenamefont{Spoden, Zinner,
  Herschbach, van Drunen, Ertmer, and Birkl}}]{Spoden:05}
\bibinfo{author}{\bibfnamefont{P.}~\bibnamefont{Spoden}},
  \bibinfo{author}{\bibfnamefont{M.}~\bibnamefont{Zinner}},
  \bibinfo{author}{\bibfnamefont{N.}~\bibnamefont{Herschbach}},
  \bibinfo{author}{\bibfnamefont{W.~J.} \bibnamefont{van Drunen}},
  \bibinfo{author}{\bibfnamefont{W.}~\bibnamefont{Ertmer}}, \bibnamefont{and}
  \bibinfo{author}{\bibfnamefont{G.}~\bibnamefont{Birkl}},
  \bibinfo{journal}{Phys. Rev. Lett.} \textbf{\bibinfo{volume}{94}},
  \bibinfo{pages}{223201} (\bibinfo{year}{2005}).

\bibitem[{\citenamefont{DeMarco et~al.}(2001)\citenamefont{DeMarco, Papp, and
  Jin}}]{DeMarco:01}
\bibinfo{author}{\bibfnamefont{B.}~\bibnamefont{DeMarco}},
  \bibinfo{author}{\bibfnamefont{S.~B.} \bibnamefont{Papp}}, \bibnamefont{and}
  \bibinfo{author}{\bibfnamefont{D.~S.} \bibnamefont{Jin}},
  \bibinfo{journal}{Phys. Rev. Lett} \textbf{\bibinfo{volume}{86}},
  \bibinfo{pages}{5409} (\bibinfo{year}{2001}).

\bibitem[{\citenamefont{Schmidt et~al.}(2003)\citenamefont{Schmidt, Hensler,
  Werner, Binhammer, G\"{o}rlitz, and Pfau}}]{Schmidt:03}
\bibinfo{author}{\bibfnamefont{P.~O.} \bibnamefont{Schmidt}},
  \bibinfo{author}{\bibfnamefont{S.}~\bibnamefont{Hensler}},
  \bibinfo{author}{\bibfnamefont{J.}~\bibnamefont{Werner}},
  \bibinfo{author}{\bibfnamefont{T.}~\bibnamefont{Binhammer}},
  \bibinfo{author}{\bibfnamefont{A.}~\bibnamefont{G\"{o}rlitz}},
  \bibnamefont{and} \bibinfo{author}{\bibfnamefont{T.}~\bibnamefont{Pfau}},
  \bibinfo{journal}{J. Opt. Soc. Am. B} \textbf{\bibinfo{volume}{20}},
  \bibinfo{pages}{960} (\bibinfo{year}{2003}).

\bibitem[{\citenamefont{Feldker et~al.}(2011)\citenamefont{Feldker, Sch\"{u}tz,
  John, and Birkl}}]{Feldker:11}
\bibinfo{author}{\bibfnamefont{T.}~\bibnamefont{Feldker}},
  \bibinfo{author}{\bibfnamefont{J.}~\bibnamefont{Sch\"{u}tz}},
  \bibinfo{author}{\bibfnamefont{H.}~\bibnamefont{John}}, \bibnamefont{and}
  \bibinfo{author}{\bibfnamefont{G.}~\bibnamefont{Birkl}},
  \bibinfo{journal}{Eur. Phys. J. D} \textbf{\bibinfo{volume}{65}},
  \bibinfo{pages}{257} (\bibinfo{year}{2011}).

\bibitem[{not({\natexlab{b}})}]{note2}
\bibinfo{note}{We do not expect a change in the rate coefficients for a
  reversal of the role of the two isotopes, although such a dependence has been
  observed for dual-species alkali mixtures as reported in J. P. Shaffer, W.
  Chalupczak, and N.P. Bigelow, Phys. Rev. A {\bf 60}, R3365 (1999).}

\bibitem[{\citenamefont{S\"{o}ding et~al.}(1998)\citenamefont{S\"{o}ding,
  Gu\'{e}ry-Odelin, Desbiolles, Ferrari, and Dalibard}}]{Soeding:98}
\bibinfo{author}{\bibfnamefont{J.}~\bibnamefont{S\"{o}ding}},
  \bibinfo{author}{\bibfnamefont{D.}~\bibnamefont{Gu\'{e}ry-Odelin}},
  \bibinfo{author}{\bibfnamefont{P.}~\bibnamefont{Desbiolles}},
  \bibinfo{author}{\bibfnamefont{G.}~\bibnamefont{Ferrari}}, \bibnamefont{and}
  \bibinfo{author}{\bibfnamefont{J.}~\bibnamefont{Dalibard}},
  \bibinfo{journal}{Phys. Rev. Lett} \textbf{\bibinfo{volume}{80}},
  \bibinfo{pages}{1869} (\bibinfo{year}{1998}).

\end{thebibliography}

\end{document}